# A Comparative Study of LOWESS and RBF Approximations for Visualization

Michal Smolik[1], Vaclav Skala[1] and Ondrej Nedved[1]

[1] Faculty of Applied Sciences, University of West Bohemia,
Univerzitni 8, CZ 30614 Plzen, Czech Republic

**Abstract.** Approximation methods are widely used in many fields and many techniques have been published already. This comparative study presents a comparison of LOWESS (Locally weighted scatterplot smoothing) and RBF (Radial Basis Functions) approximation methods on noisy data as they use different approaches. The RBF approach is generally convenient for high dimensional scattered data sets. The LOWESS method needs finding a subset of nearest points if data are scattered. The experiments proved that LOWESS approximation gives slightly better results than RBF in the case of lower dimension, while in the higher dimensional case with scattered data the RBF method has lower computational complexity.

**Keywords:** Radial Basis Functions, LOWESS, Approximation

Notation used

$D$: dimension
$K$: $k$-nearest points
$M$: number of radial basis functions for approximation
$N$: number of all input points
$R$: number of points at which the approximation is calculated
$\xi$: point where to calculate the approximation
$d$: degree of polynomial
$r$: $r = d + 2$
$q$: $q = d + 1$

## 1 Introduction

Interpolation and approximation techniques are often used in data processing. Approximation methods of values $\boldsymbol{y}_i$ in the given $\{(\boldsymbol{x}_i, \boldsymbol{y}_i)\}_1^N$ data set lead to a smooth function which minimizes the difference between given data and the determined function [13]. It can be used for visualization of noisy data [1, 2], visualization of the basic shape of measured/calculated data [9], for prediction, and other purposes. Many methods have been described together with their properties. This paper describes LOWESS (Locally weighted scatterplot smoothing) and RBF (Radial basis functions) methods and their experimental comparison.

## 2 LOWESS

The locally weighted scatterplot smoothing method (LOWESS) [3] is often used, especially in statistical applications. The value of an approximated function at a point $x_0$ is calculated from the formula of a curve which minimizes a sum $S$ in the $k$-nearest neighborhood (KNN) points of the given point $\xi$.

$$S = \sum_{i=1}^{K} \omega_i \cdot \left(y_i - P_{(d)}(x_i)\right)^2, \tag{1}$$

where $P_{(d)}(x) = a_0 + a_1 x + a_2 x^2 + \ldots + a_d x^d$ is a $d$ degree of a polynomial function with unknown coefficients $\boldsymbol{a} = [a_0, a_1, a_2, \ldots, a_d]^T$. We can rewrite the sum $S$ in a matrix form as:

$$S = (\boldsymbol{b} - \boldsymbol{Aa})^T \cdot \boldsymbol{W} \cdot (\boldsymbol{b} - \boldsymbol{Aa}), \tag{2}$$

where $\boldsymbol{b} = [y_1, y_2, \ldots, y_K]^T$ is a vector of function values, matrix $\boldsymbol{A}$ is equal to:

$$\boldsymbol{A} = \begin{bmatrix} 1 & x_1 & \cdots & x_1^d \\ 1 & x_2 & \cdots & x_2^d \\ \vdots & & \ddots & \vdots \\ 1 & x_K & \cdots & x_K^d \end{bmatrix} \tag{3}$$

and matrix $\boldsymbol{W}$ is a diagonal matrix:

$$\boldsymbol{W} = \begin{bmatrix} \omega(\|x_1 - \xi\|) & & & 0 \\ & \omega(\|x_2 - \xi\|) & & \\ & & \ddots & \\ 0 & & & \omega(\|x_K - \xi\|) \end{bmatrix} = \begin{bmatrix} \omega_1 & & & 0 \\ & \omega_2 & & \\ & & \ddots & \\ 0 & & & \omega_K \end{bmatrix}, \tag{4}$$

where $\omega(r)$ are weighting functions, which have to satisfy the following conditions defined as:

$$\forall a, b \in [0; 1], a < b : \omega(a) \geq \omega(b) \ \wedge \ \omega(0) = 1 \ \wedge \forall c \geq 1 : \omega(c) = 0. \tag{5}$$

One such example of a weighting function $\omega$ can be the tricube function:

$$\omega(r = \|x_i - \xi\|) = \omega_i = \begin{cases} (1 - r^3)^3 & r \in \langle 0; 1 \rangle \\ 0 & r > 1 \end{cases}. \tag{6}$$

Equation (2) can be modified as:

$$\begin{aligned} S &= \boldsymbol{b}^T \boldsymbol{W} \boldsymbol{b} - \boldsymbol{b}^T \boldsymbol{W} \boldsymbol{Aa} - (\boldsymbol{Aa})^T \boldsymbol{W} \boldsymbol{b} + (\boldsymbol{Aa})^T \boldsymbol{W} \boldsymbol{Aa} \\ &= \boldsymbol{b}^T \boldsymbol{W} \boldsymbol{b} - \boldsymbol{b}^T \boldsymbol{W} \boldsymbol{Aa} - \boldsymbol{a}^T \boldsymbol{A}^T \boldsymbol{W} \boldsymbol{b} + \boldsymbol{a}^T \boldsymbol{A}^T \boldsymbol{W} \boldsymbol{Aa}. \end{aligned} \tag{7}$$

The sum $S$ is minimal if the partial derivative of $S$ with respect to $\boldsymbol{a}$ is equal to zero:

$$\frac{\partial S}{\partial \boldsymbol{a}} = -(\boldsymbol{b}^T \boldsymbol{W} \boldsymbol{A})^T - \boldsymbol{A}^T \boldsymbol{W} \boldsymbol{b} + 2\boldsymbol{A}^T \boldsymbol{W} \boldsymbol{Aa} = \boldsymbol{0} \tag{8}$$

as $\boldsymbol{W} = \boldsymbol{W}^T$ and therefore:

$$A^T W A a = A^T W b$$
$$a = (A^T W A)^{-1} A^T W b. \tag{9}$$

The numerical stability of calculations is influenced by the position of the interval of the $k$-nearest neighborhood points of the point $\xi$. The LOWESS approximation is "locally" based, as only $k$-nearest points are used and thus $r$ is actually computed as $r = \|x_i - \xi\|$. To solve problems with the numerical stability of calculations and independence of absolute position, we have to use relative position of all the $k$-nearest neighborhood points of the point $\xi$ such that the matrix $A$ from (3) is defined as:

$$A = \begin{bmatrix} 1 & (x_1 - \xi) & \cdots & (x_1 - \xi)^d \\ 1 & (x_2 - \xi) & \cdots & (x_2 - \xi)^d \\ \vdots & & \ddots & \vdots \\ 1 & (x_K - \xi) & \cdots & (x_K - \xi)^d \end{bmatrix} \tag{10}$$

## 2.1 LOWESS with linear regression

Linear regression, i.e. choosing $d = 1$, appears to strike a good balance between computational simplicity and the flexibility needed to reproduce patterns in the data. In such a case, we can rewrite (9) as:

$$a = \begin{bmatrix} \sum_{i=1}^{K} \omega_i & \sum_{i=1}^{K} \omega_i x_i \\ \sum_{i=1}^{K} \omega_i x_i & \sum_{i=1}^{K} \omega_i x_i^2 \end{bmatrix}^{-1} \cdot \begin{bmatrix} \sum_{i=1}^{K} \omega_i y_i \\ \sum_{i=1}^{K} \omega_i x_i y_i \end{bmatrix} \tag{11}$$

and after some adjustments we can get a final formula for unknown coefficients $a$:

$$\begin{bmatrix} a_0 \\ a_1 \end{bmatrix} = \frac{1}{(\sum_{i=1}^{K} \omega_i) \cdot (\sum_{i=1}^{K} \omega_i x_i^2) - (\sum_{i=1}^{K} \omega_i x_i)^2}$$
$$\cdot \begin{bmatrix} \left( \sum_{i=1}^{K} \omega_i y_i \right) \left( \sum_{i=1}^{K} \omega_i x_i^2 \right) - \left( \sum_{i=1}^{K} \omega_i x_i \right) \left( \sum_{i=1}^{K} \omega_i x_i y_i \right) \\ -\left( \sum_{i=1}^{K} \omega_i y_i \right) \left( \sum_{i=1}^{K} \omega_i x_i \right) + \left( \sum_{i=1}^{K} \omega_i \right) \left( \sum_{i=1}^{K} \omega_i x_i y_i \right) \end{bmatrix} \tag{12}$$

## 2.2 LOWESS with constant regression

Constant regression, i.e. choosing $d = 0$, is the most computationally simple, but from a practical point of view, an assumption of local linearity seems to serve far better than an assumption of local constancy because the tendency is to plot variables that are related to one another. Thus, the linear LOWESS regression produces better results than the constant LOWESS regression, which is very simple. In this case, we can rewrite it from (9) as:

$$a_0 = \frac{\sum_{i=1}^{K} \omega_i y_i}{\sum_{i=1}^{K} \omega_i}. \tag{13}$$

Comparing formulas from (13) and (12), it can be seen that LOWESS with constant regression is computationally much easier than LOWESS with linear regression.

## 3 Radial Basis Functions

Radial basis functions (RBF) [4, 11, 12] is based on distances, generally in $D$-dimensional space. The value of an approximated function at a point $x$ is calculated from the formula:

$$f(x) = \sum_{i=1}^{M} \lambda_i \Phi(\|x - \xi_i\|) + P_d(x), \tag{14}$$

where $P_{(d)}(x) = a_0 + a_1 x + a_2 x^2 + \ldots + a_d x^d$ is a $d$ degree polynomial function with unknown coefficients $\boldsymbol{a} = [a_0, a_1, a_2, \ldots, a_d]^T$, $M$ is the number of radial basis functions, and $\boldsymbol{\lambda} = [\lambda_1, \ldots, \lambda_M]$ are weights of radial basis functions $\Phi(\|x - \xi_i\|)$. The function $\Phi$ is a real-valued function whose value depends only on the distance from some other point $\xi_i$, called a center, so that:

$$\Phi_i(x) = \Phi(\|x - \xi_i\|). \tag{15}$$

As the values $f(x_i)$ at a point $x_i$ are known, equation (14) represents a system of linear equations that has to be solved in order to determine coefficients $\boldsymbol{\lambda}$ and $\boldsymbol{a}$, i.e.

$$f(x_j) = \sum_{i=1}^{M} \lambda_i \Phi(\|x_j - \xi_i\|) + P_d(x_j) \text{ for } \forall j \in \{1, \ldots, N\}. \tag{16}$$

Using matrix notation we can rewrite (16) as:

$$\begin{bmatrix} \Phi(\|x_1 - \xi_1\|) & \cdots & \Phi(\|x_1 - \xi_M\|) & 1 & x_1 & \cdots & x_1^d \\ \vdots & & \vdots & \vdots & \vdots & & \vdots \\ \Phi(\|x_N - \xi_1\|) & \cdots & \Phi(\|x_N - \xi_M\|) & 1 & x_N & \cdots & x_N^d \end{bmatrix} \cdot \begin{bmatrix} \lambda_1 \\ \vdots \\ \lambda_M \\ a_0 \\ \vdots \\ a_d \end{bmatrix} = \begin{bmatrix} f(x_1) \\ \vdots \\ f(x_N) \end{bmatrix}. \tag{17}$$

We can create a "simple" RBF formula, see (18), using (17) with only one radial basis function, i.e. $M = 1$. This formula can be used in the same manner as the LOWESS method for calculating approximated value at the point $\xi$, using only the $k$-nearest neighborhood points of the point $\xi$, which is the center of radial basis function $\phi(\|x - \xi\|)$, too.

$$\begin{bmatrix} \Phi(\|x_1 - \xi\|) & 1 & x_1 & \cdots & x_1^d \\ \vdots & \vdots & \vdots & & \vdots \\ \Phi(\|x_K - \xi\|) & 1 & x_K & \cdots & x_K^d \end{bmatrix} \cdot \begin{bmatrix} \lambda_1 \\ a_0 \\ \vdots \\ a_d \end{bmatrix} = \begin{bmatrix} f(x_1) \\ \vdots \\ f(x_K) \end{bmatrix} \rightarrow \boldsymbol{A} \cdot \boldsymbol{\lambda} = \boldsymbol{f} . \quad (18)$$

The coefficients $\boldsymbol{\eta} = [\lambda_1, \boldsymbol{a}^T]^T$ in overdetermined system of linear equations (18) are computed by the least squares error method:

$$\boldsymbol{\eta} = (A^T A)^{-1} \cdot (A^T \boldsymbol{f}) . \quad (19)$$

As the numerical stability of calculations is influenced by the position of the interval of the $k$-nearest neighborhood points of the point $\xi$ and the RBF approximation is "locally" based, only $k$-nearest points are used. To solve problems with the numerical stability of calculations, we have to move all the $k$-nearest neighborhood points of the point $\xi$ such that the matrix $\boldsymbol{A}$ from (18) is defined as:

$$\boldsymbol{A} = \begin{bmatrix} \Phi(\|x_1 - \xi\|) & 1 & (x_1 - \xi) & \cdots & (x_1 - \xi)^d \\ \vdots & \vdots & \vdots & & \vdots \\ \Phi(\|x_K - \xi\|) & 1 & (x_K - \xi) & \cdots & (x_K - \xi)^d \end{bmatrix} \quad (20)$$

and $f(x)$ is defined as:

$$f(x) = \lambda_1 \Phi(\|x - \xi\|) + P_d(x - \xi) . \quad (21)$$

For locally-based approximation, any compactly supported radial basis function (CSRBF) [8, 12] can be used. CSRBF is a function defined on $r \in \langle 0; 1 \rangle$, is equal to 0 for all $r > 1$, and has to satisfy the conditions in (5). In the tests presented here, the $\Phi(r)$ function was selected as:

$$\Phi(r) = \begin{cases} (1 - r^3)^3 & r \in \langle 0; 1 \rangle \\ 0 & r > 1 \end{cases} , \quad (22)$$

which is exactly the same function as weighting function (6) for LOWESS approximation.

### 3.1 Simplified RBF with a constant polynomial

Choosing $d = 0$, we will get a polynomial of zero degree which is only a constant, i.e.; $P_d = a_0$.

$$\begin{bmatrix} \Phi(\|x_1 - \xi\|) & 1 \\ \vdots & \vdots \\ \Phi(\|x_K - \xi\|) & 1 \end{bmatrix} \cdot \begin{bmatrix} \lambda_1 \\ a_0 \end{bmatrix} = \begin{bmatrix} f(x_1) \\ \vdots \\ f(x_K) \end{bmatrix} \rightarrow \boldsymbol{A} \cdot \boldsymbol{\eta} = \boldsymbol{f} . \quad (23)$$

It leads to overdetermined system of linear equations. Using the method of least squares, we can calculate $\boldsymbol{\eta}$:

$$\boldsymbol{\eta} = (A^T A)^{-1} \cdot (A^T \boldsymbol{f}) , \quad (24)$$

where $\boldsymbol{\eta} = [\lambda_1, a_0]^T$.

$$\begin{bmatrix} \lambda_1 \\ a_0 \end{bmatrix} = \begin{bmatrix} \sum_{i=1}^{K} \left( \Phi(\|x_i - \xi\|) \right)^2 & \sum_{i=1}^{K} \Phi(\|x_i - \xi\|) \\ \sum_{i=1}^{K} \Phi(\|x_i - \xi\|) & \sum_{i=1}^{K} 1 \end{bmatrix}^{-1} \cdot \begin{bmatrix} \sum_{i=1}^{K} \Phi(\|x_i - \xi\|) \cdot f(x_i) \\ \sum_{i=1}^{K} f(x_i) \end{bmatrix}, \quad (25)$$

where $\sum_{i=1}^{K} 1 = K$ and after adjustments:

$$\begin{bmatrix} \lambda_1 \\ a_0 \end{bmatrix} = \frac{1}{\left( \sum_{i=1}^{K} \left( \Phi(\|x_i - \xi\|) \right)^2 \right) \cdot K - \left( \sum_{i=1}^{K} \Phi(\|x_i - \xi\|) \right)^2}$$

$$\cdot \begin{bmatrix} K & -\sum_{i=1}^{K} \Phi(\|x_i - \xi\|) \\ -\sum_{i=1}^{K} \Phi(\|x_i - \xi\|) & \sum_{i=1}^{K} \left( \Phi(\|x_i - \xi\|) \right)^2 \end{bmatrix} \begin{bmatrix} \sum_{i=1}^{K} \Phi(\|x_i - \xi\|) \cdot f(x_i) \\ \sum_{i=1}^{K} f(x_i) \end{bmatrix}. \quad (26)$$

The value $f(\xi)$ is calculated as:

$$f(\xi) = \lambda_1 \Phi(\|\xi - \xi\|) + a_1 = \lambda_1 \Phi(0) + a_0. \quad (27)$$

### 3.2 Simplified RBF without a polynomial

In the case of using simplified RBF without polynomial $P_d$, we get the following equation:

$$\begin{bmatrix} \Phi(\|x_1 - \xi\|) \\ \vdots \\ \Phi(\|x_K - \xi\|) \end{bmatrix} \cdot [\lambda_1] = \begin{bmatrix} f(x_1) \\ \vdots \\ f(x_K) \end{bmatrix} \rightarrow \boldsymbol{A} \cdot \lambda_1 = \boldsymbol{f}, \quad (28)$$

where $\boldsymbol{A}$ and $\boldsymbol{f}$ are column vectors. Using the method of least squares, we can calculate $\lambda_1$:

$$\lambda_1 = \frac{\boldsymbol{A}^T \cdot \boldsymbol{f}}{\boldsymbol{A}^T \boldsymbol{A}}. \quad (29)$$

Equation (29) can be rewritten as:

$$\lambda_1 = \frac{\sum_{i=1}^{K} \Phi(\|x_i - \xi\|) \cdot f(x_i)}{\sum_{i=1}^{K} \left( \Phi(\|x_i - \xi\|) \right)^2}. \quad (30)$$

The value $f(\xi)$ is calculated as:

$$f(\xi) = \lambda_1 \Phi(\|\xi - \xi\|) = \lambda_1 \Phi(0). \quad (31)$$

# 4 Comparison of Time Complexity

In the following, a comparison of LOWESS and RBF will be made. The main criteria for comparison are:

- The computational complexity, which is critical if many points have to be approximated.
- The quality of the final approximation (see section 5).

## 4.1 LOWESS

The size of matrix $\boldsymbol{A}$ is $k \times q$, where the number of used nearest points is $k$ and $q$ is equal to the degree of the polynomial plus 1. The size of diagonal matrix $\boldsymbol{W}$ is $k \times k$, the size of vector $\boldsymbol{b}$ is $k \times 1$ and the size of vector $\mathbf{x}$ is $k \times 1$. The time complexity of LOWESS using equation (9) can be calculated in the following way:

$$
\begin{array}{lcl}
\boldsymbol{A}^T \boldsymbol{W} \boldsymbol{A} & \rightarrow & O(q^2 k + qk) \\
(\boldsymbol{A}^T \boldsymbol{W} \boldsymbol{A})^{-1} & \rightarrow & O(q^2 k + qk + q^3) \\
\boldsymbol{A}^T \boldsymbol{W} \boldsymbol{b} & \rightarrow & O(2qk) \\
(\boldsymbol{A}^T \boldsymbol{W} \boldsymbol{A})^{-1} \boldsymbol{A}^T \boldsymbol{W} \boldsymbol{b} & \rightarrow & O(k(q^2 + 3q) + q^3 + q^2)
\end{array}
\tag{32}
$$

As the size $k$ of matrix $\boldsymbol{A}$ is much larger than the size $q$ of matrix $\boldsymbol{A}$, the time complexity from (32) will become:

$$
\begin{array}{lll}
O(3qk) & for & q = \{1,2\} \\
O(q^2 k) & for & q \geq 3
\end{array}
\tag{33}
$$

The time complexity of LOWESS when calculating the approximation value in $R$ points will become:

$$
\begin{array}{lll}
O(N \log N + R \cdot 3qk) & for & q = \{1,2\} \\
O(N \log N + R \cdot q^2 k) & for & q \geq 3
\end{array}
\tag{34}
$$

where $N$ is the number of input points and $O(N \log N)$ is the time complexity of the sorting algorithm for 1&½ dimensional data. In the case of higher dimensions $D$&½, i.e. $D > 1$, the total time complexity of selecting $k$-nearest points from $N$ points increases (see section 7 for more details).

## 4.2 Simplified RBF

The size of matrix $\boldsymbol{A}$ is $k \times r$, where the number of used nearest points is $k$ and $r$ is equal to the degree of the polynomial plus 2. The size of vector $\boldsymbol{f}$ is $k \times 1$ and the size of vector $\boldsymbol{\eta} = [\boldsymbol{\lambda}^T, \boldsymbol{a}^T]^T$ is $k \times 1$. The time complexity of RBF using equation (24) can be calculated in the following way:

$$
\begin{array}{rcc}
\boldsymbol{A}^T\boldsymbol{A} & \rightarrow & O(r^2 k) \\
(\boldsymbol{A}^T\boldsymbol{A})^{-1} & \rightarrow & O(r^2 k + r^3) \\
\boldsymbol{A}^T\boldsymbol{f} & \rightarrow & O(rk) \\
(\boldsymbol{A}^T\boldsymbol{A})^{-1}(\boldsymbol{A}^T\boldsymbol{f}) & \rightarrow & O(k(r^2 + r) + r^3 + r^2)
\end{array}
\tag{35}
$$

As the size $k$ of matrix $\boldsymbol{A}$ is much larger than the size $r$ of matrix $\boldsymbol{A}$, the time complexity from (35) will become:

$$
O(r^2 k) \tag{36}
$$

The time complexity of simplified RBF when calculating the approximation value in $R$ points can be estimated:

$$
O(N \log N + R \cdot r^2 k) \tag{37}
$$

where $N$ is the number of input points and $O(N \log N)$ is the time complexity of the sorting algorithm for 1&½ dimensional data.

## 5    Comparison of Measured Errors

For a demonstration of LOWESS and RBF approximation properties, the standard testing function, which is considered by Hickernell and Hon [5], has been selected:

$$
\tau(x) = e^{\left[-15\left(\left(x-\frac{1}{2}\right)^2\right)\right]} + \frac{1}{2}e^{\left[-20\left(\left(x-\frac{1}{2}\right)^2\right)\right]} - \frac{3}{4}e^{\left[-8\left(\left(x+\frac{1}{2}\right)^2\right)\right]} \tag{38}
$$

This function was sampled at $\langle -1,1 \rangle$. We added random noise with uniform distribution from interval $\langle -0.1, 0.1 \rangle$ and used it as input for both methods. The following graphs present the behavior of the LOWESS and RBF approximations.

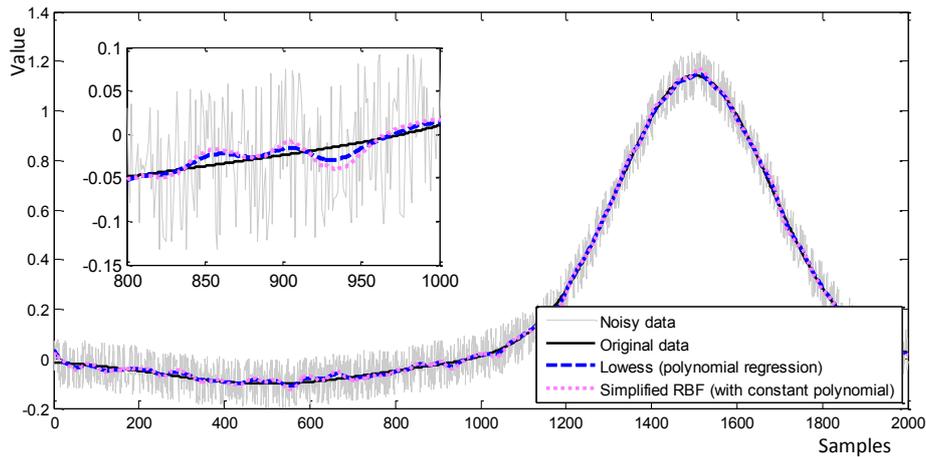

Graph 1: Comparison of LOWESS with Simplified RBF. 100 nearest samples out of 2000 total were used as values for approximation; sampled interval: $\langle -1,1 \rangle$.

Function values are on the vertical axis; values on the horizontal axis are sample indices. Since the function is sampled at $\langle-1,1\rangle$ and the number of samples is 2000, the sampling rate is 1000 samples per 1 unit.

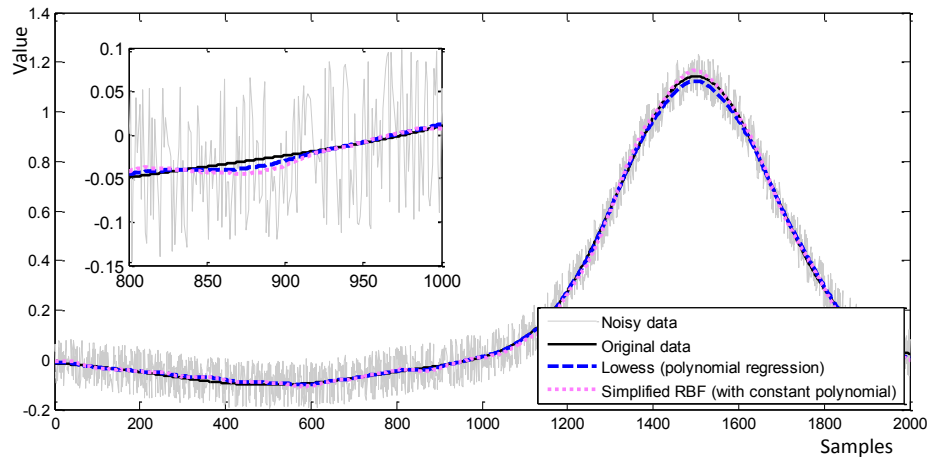

Graph 2: Comparison of LOWESS with Simplified RBF. 200 nearest samples out of 2000 total were used as values for approximation; sampled interval: $\langle-1,1\rangle$.

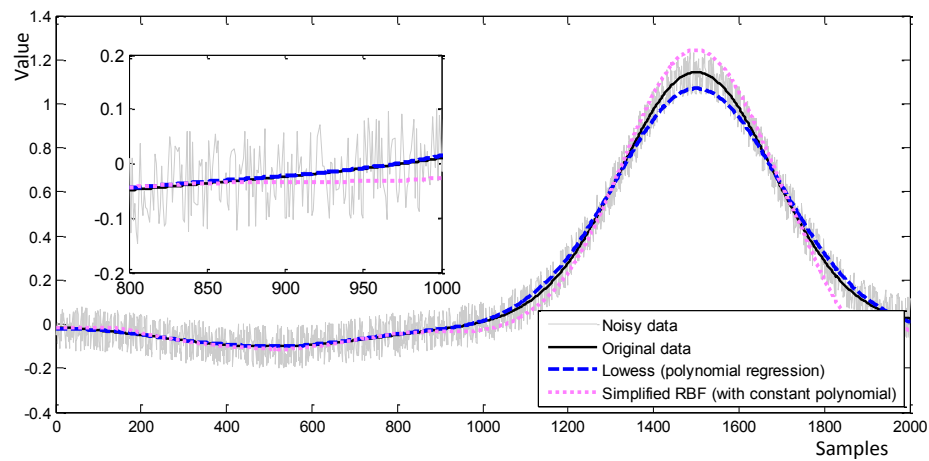

Graph 3: Comparison of LOWESS with Simplified RBF. 500 nearest samples out of 2000 total were used as values for approximation; sampled interval: $\langle-1,1\rangle$.

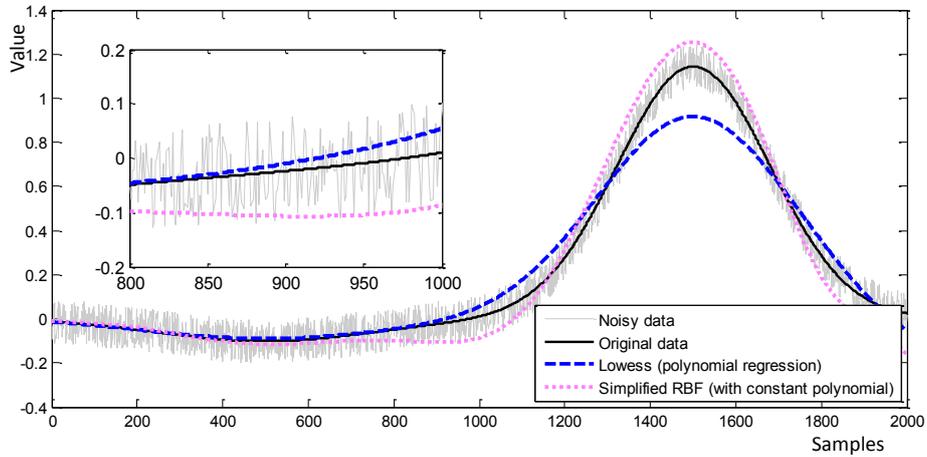

Graph 4: Comparison of LOWESS with Simplified RBF. 1000 nearest samples out of 2000 total were used as values for approximation; sampled interval: $\langle -1,1 \rangle$.

The error of approximation can be measured in different ways. The first one is to measure the change of the first derivative, which is the curvature of the resulting curve. If the first derivative changes too much, then the curve is jagged; on the contrary, if the first derivative does not change too much, then the curve is smooth. The absolute error can be calculated using the formula:

$$E_c = \sum_{i=1}^{N} \|f''(x_i)\|, \tag{39}$$

where $f''(x_i)$ is calculated using the formula:

$$f''(x_i) = \frac{f(x_{i+1} - 2x_i + x_{i-1})}{(x_{i+1} - x_i)(x_i - x_{i-1})}, \tag{40}$$

Let $\boldsymbol{p} = [x, f(x)]$ be the approximated point in current space ($2D$ for $1\&\frac{1}{2}$ dimensions) and $\boldsymbol{\kappa} = [x, \tau(x)]$ be a point of the sampled function (38), which is a set $K = \{\boldsymbol{\kappa}_1, \dots, \boldsymbol{\kappa}_N\} = \{[x_1, \tau(x_1)], \dots, [x_N, \tau(x_N)]\}$, then the distance error from the original curve without noise can be calculated as:

$$E_d = \sum_{i=1}^{N} \|\boldsymbol{p}_i - \boldsymbol{\xi}_j\|, \text{where } \|\boldsymbol{p}_i - \boldsymbol{\kappa}_j\| \text{ is minimal } \forall j \in \{1, \dots, N\} \text{ for given } i. \tag{41}$$

Let us note that the distance is not measured vertically to the curve but "orthogonally" to the curve. Using formulas (39) and (41), we can show the following table of calculated errors.

Tab. 1: Measured errors for graphs Graph 1 - Graph 4 (for $N = 2000$).

| $k$-nearest samples | $E_c$ | | $E_d$ | |
|---|---|---|---|---|
| | LOWESS | Simplified RBF | LOWESS | Simplified RBF |
| 100 | 0.0721 | 1.4585 | 7.2997 | 12.5647 |
| 200 | 0.0212 | 0.7689 | 10.5378 | 14.7898 |
| 500 | 0.0132 | 0.3103 | 15.6759 | 40.5985 |
| 1 000 | 0.0091 | 0.1618 | 45.0717 | 70.8979 |

Some comparison results can be seen using (Tab. 1). The LOWESS approximation is always smoother (according to measured error $E_c$) and closer to the original data without noise (according to measured error $E_d$) when using the same $k$-nearest samples.

## 6 Global RBF Approximation

Global RBF approximation can be calculated using (14). In this case, the whole data set has to be processed at once. Compared to the simplified version of RBF approximation, we only get one $\boldsymbol{\lambda}$ vector for all input samples and thus we solve a linear system only once. Moreover, we do not need to sort the input points in any way, unlike LOWESS and simplified RBF approximations, which were presented in previous sections. The global RBF approximation is calculated using the following formula (from (17)):

$$A\boldsymbol{\lambda} = \boldsymbol{f} \rightarrow \boldsymbol{\lambda} = (A^T A)^{-1} \cdot (A^T \boldsymbol{f}) \tag{42}$$

where the size of matrix $\boldsymbol{A}$ is $N \times (M + d + 1)$, $N$ is the number of input points, $M$ is the number of radial basis functions, $d$ is the degree of the polynomial, the size of vector $\boldsymbol{f}$ is $N \times 1$, the size of vector $\boldsymbol{\lambda} = [\lambda_1, \dots, \lambda_M, a_0, \dots, a_d]^T$ is $(M + d + 1) \times 1$. We can express the time complexity of the global RBF approximation calculation (42) as:

$$
\begin{array}{lcc}
A^T A & \rightarrow & O((M + d + 1)^2 N) \\
(A^T A)^{-1} & \rightarrow & O((M + d + 1)^2 N + (M + d + 1)^3) \\
A^T \boldsymbol{f} & \rightarrow & O\big((M + d + 1)N\big) \\
(A^T A)^{-1} \cdot (A^T \boldsymbol{f}) & \rightarrow & O\left(\begin{array}{c} N\big((M + d + 1)^2 + (M + d + 1)\big) + \\ +(M + d + 1)^3 + (M + d + 1)^2 \end{array}\right)
\end{array} \tag{43}
$$

and after leaving only the most complex part, the time complexity is:

$$O(M^2 N) \tag{44}$$

We sampled function (38), added random noise with uniform distribution from interval $\langle -0.1, 0.1 \rangle$, and used that data as input for both methods mentioned in previous sections (LOWESS and simplified RBF) and for global RBF approximation as well. The following graph presents the behavior of the LOWESS, simplified RBF and global RBF approximations. Function values are on the vertical axis, values on the horizontal axis are sample indices. Since the function is sampled at ⟨−1,1⟩ and the number of samples is 2000, the sampling rate is 1000 samples per 1 unit.

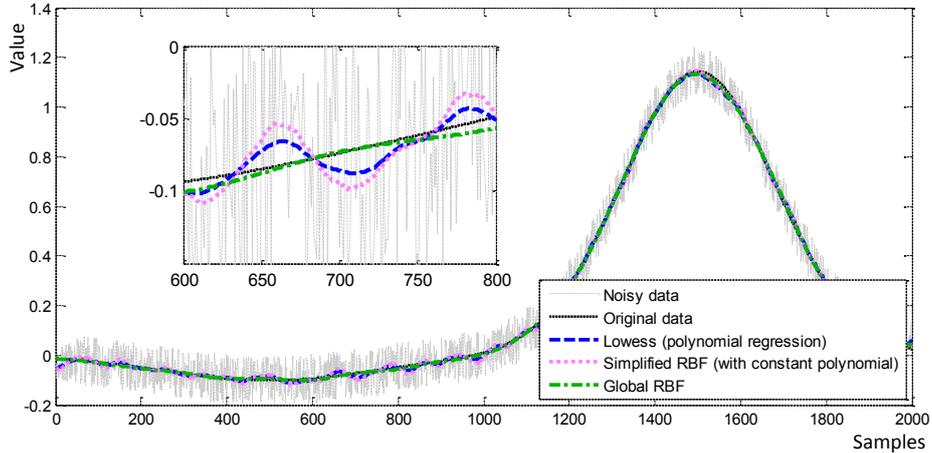

Graph 5: Comparison of LOWESS and Simplified RBF with global RBF. 100 nearest samples out of 2000 total were used as values for local approximation, which gives 20 pivots (lambdas) for global RBF; sampled interval: ⟨−1,1⟩.

The following table presents calculated errors using formulas (39) and (41) (for Graph 5) for all approximation methods described in this paper.

Tab. 2: Measured errors for Graph 5.

| $E_c$ | | | $E_d$ | | |
|---|---|---|---|---|---|
| LOWESS | Simplified RBF | Global RBF | LOWESS | Simplified RBF | Global RBF |
| 0.0718 | 1.5266 | 0.0168 | 10.6785 | 16.0734 | 6.0123 |

It can be seen, that global RBF approximation is closer to the original data and smoother than simple RBF or even LOWESS approximation. For the situation in Graph 5 and Tab. 2, the time complexity of global RBF is exactly the same as the time complexity of both other methods when calculating the approximation at all input points.

# 7 Approximation in Higher Dimensions

Let as assume that a scattered data approximation [6, 7] in 2&½ or 3&½ dimensions have to be made, i.e. $D$&½ dimensions have to be made. In the following, we describe the expansion of LOWESS and RBF approximation algorithms into higher dimensions.

In higher than 1&½ dimensions, we have to deal with the fact that there is no ordering defined in general. Thus, we cannot sort all input points at once in the beginning and then choose $k$-nearest points with $O(1)$ time complexity. The time complexity of selecting $k$-nearest points from $N$ points is $O(N \log N)$, and thus the time complexity of LOWESS or Simple RBF approximation can be estimated as:

$$O\left(R \cdot \left(N \log N + \left\{\begin{matrix} O_{LOWESS} \\ or \\ O_{RBF} \end{matrix}\right\}\right)\right), \tag{45}$$

where $O_{LOWESS}$ is the same time complexity as the time complexity of LOWESS approximation in 1&½ dimensions and $O_{RBF}$ is the same time complexity as the time complexity of Simple RBF approximation in 1&½ dimensions.

## 7.1 LOWESS

In the case of $D$&½ dimensional approximation, we have to change the notation in (1) as $\boldsymbol{x}$ is a $D$-dimensional position vector:

$$S = \sum_{i=1}^{N} \omega_i \cdot \left(h_i - P^{(D)}(\boldsymbol{x_i})\right)^2, \tag{46}$$

where $h = P^{(D)}(\boldsymbol{x})$ is a $D$ dimensional hypersurface function with unknown coefficients $\boldsymbol{a} = [a_0, a_1, a_2, \dots, a_k]^T$. For $D = 2$, we can write $P^{(D)}(\boldsymbol{x})$, for example, like:

$$P^{(2)}(\boldsymbol{x}) = a_0 + a_1 x + a_2 y + a_3 x^2 + a_4 y^2 + a_5 xy, \tag{47}$$

where $\boldsymbol{x} = [x, y]^T$. The matrix $\boldsymbol{A}$ is then equal to:

$$\boldsymbol{A} = \begin{bmatrix} 1 & x_1 & y_1 & x_1^2 & y_1^2 & x_1 y_1 \\ 1 & x_2 & y_2 & x_2^2 & y_2^2 & x_2 y_2 \\ \vdots & \vdots & \vdots & \vdots & \vdots & \vdots \\ 1 & x_N & y_N & x_N^2 & y_N^2 & x_N y_N \end{bmatrix} \tag{48}$$

We can omit some coefficients $a_i$ and corresponding columns in matrix $\boldsymbol{A}$, where $i \in \{0, 1, \dots, 5\}$. All other computations remain the same.

The computation complexity will increase as the size of matrix $\boldsymbol{A}$ increases. However, if we use a constant hypersurface function with only one coefficient $a_0$, then the time complexity does not change with different dimensions $D$.

## 7.2 Simplified RBF

The RBF approximation is formally independent from the dimension $D$. Therefore, all the computations remain the same as described above. The computation complexity increases slightly as the complexity of polynomial/hypersurface $P^{(D)}(\boldsymbol{x})$ increases. However, if we use a constant hypersurface function with only one coefficient $a_0$, then the time complexity will not change with different dimensions $D$. The polynomial $P^{(D)}(\boldsymbol{x})$ is actually a data approximation using a basic function and $\sum_{i=1}^{M} \lambda_i \Phi_i(r)$ controls the perturbation from $P^{(D)}(\boldsymbol{x})$.

## 8    Conclusion

We have introduced the LOWESS method of approximation and modified RBF approximation, which is comparable with LOWESS. Both methods use the same number of nearest samples for approximation and the time complexity of both these methods is the same. We calculated the distance of approximated noisy data to the original data. In all cases, for the same number of nearest samples for approximation, LOWESS gives better results. Another comparison of both methods is calculation of the smoothness. The LOWESS approximation gives us smoother results than the Simple RBF approximation. However, both these methods use a different approach for approximation than global RBF approximation; we compared them with global RBF approximation as well. Using global RBF approximation we can achieve better results (closer distance to original data and smoother approximation) when having the same time complexity of calculation. Moreover, we get one simple continuous formula and not only function values at discrete points. On the other hand, both methods can be used in higher dimensions, but the time complexity will increase for both of them compared to the situation in 1&½ dimensions. Due to this fact, in higher dimensions, global RBF approximation has lower time complexity than either LOWESS or Simple RBF approximation due to necessity of finding $k$-nearest neighbor points. Therefore, the global RBF approximation is recommendable for approximation of scattered data in higher dimensions, i.e. 2&½ dimensions and higher.

All methods for approximation compared in this paper were implemented and tested in MATLAB.

**Acknowledgements.** The authors would like to thank their colleagues at the University of West Bohemia, Plzen, for their comments and suggestions, and anonymous reviewers for their valuable critical comments and advice. The research was supported by MSMT CR projects LH12181 and SGS 2016-013.